\documentclass[twocolumn,PRL,showpacs,amsmath,amssymb,nofootinbib,aps,floatfix,notitlepage,preprintnumbers,useAMS,superscriptaddress]{revtex4}

\usepackage{graphicx}
\usepackage{xcolor}
\usepackage{aas_macros}

\newcommand{\be}{\begin{equation}}
\newcommand{\ee}{\end{equation}}

\newcommand{\p}{\mathcal{P}}
\newcommand{\Nopt}{N_{\rm{det}}}
\newcommand{\fp}{\tilde{\mathcal{P}}}

\newcommand{\MHI}{M_{\rm{HI}}}
\newcommand{\T}{\mathcal{T}}

\begin{document}
\title{Canceling out intensity mapping foregrounds}
\author{Patrick C. Breysse}
\affiliation{Canadian Institute for Theoretical Astrophysics, University of Toronto, 60 St. George Street, Toronto, ON, M5S 3H8, Canada}

\author{Christopher J. Anderson}
\affiliation{NASA Goddard Space Flight Center, Greenbelt, MD 20771, USA}

\author{Philippe Berger}
\affiliation{Jet Propulsion Laboratory, California Institute of Technology 4800 Oak Grove Dr, M/S 169-237, Pasadena CA 91109, USA}

\date{\today}
\label{firstpage}

\begin{abstract}
21 cm intensity mapping has arisen as a powerful probe of the high-redshift universe, but its potential is limited by extremely bright foregrounds and high source confusion. In this Letter, we propose a new analysis which can help solve both problems. From the combination of an intensity map with an overlapping galaxy survey we construct a new one-point statistic which is unbiased by foregrounds and contains information left out of conventional analyses. We show that our method can measure the HI mass function with unprecedented precision using observations similar to recent 21 cm detections.
\end{abstract}

\maketitle

Many experiments are studying the evolution of the universe with the redshifted 21 cm line \cite{Tingay2013,vanHaarlem2013,Bandura2014,Ali2015,Xu2015,Newburgh2016,DeBoer2017}.  These experiments seek to map vast swaths of the sky from the local universe to cosmic dawn using line intensity mapping \cite{Suginohara1999,Morales2010,Pritchard2012,Kovetz2017}.  Intensity maps do not resolve individual emitters, but instead map fluctuations in the density of neutral hydrogen.  This gives them sensitivity to the aggregate emission from all galaxies, as well as the neutral intergalactic medium.  By targeting a narrow emission line, maps can be made in three dimensions by observing at many closely-spaced frequencies.  Intensity mapping surveys can quickly access large volumes of space, allowing unprecedented constraints on cosmology and fundamental physics \cite{Shaw2015,Liu2016,EwallWice2016,Liu2016b,Foreman2018}.  In addition, they can provide statistical information about faint objects below detection thresholds of conventional surveys.  Theory predicts that the number of detected objects is dwarfed by these systems, which should then accrete and merge to form galaxies such as our Milky Way \cite{Somerville2015}.

Currently, the largest limiting factor in 21 cm cosmology is the presence of foreground emission which is typically orders of magnitude brighter than the signal \cite{Oh2003,Wang2006,Liu2011}.  These foregrounds have limited any attempts to observe HI in autocorrelation.  There have, however, been detections of cross-spectra between 21 maps and optical galaxy surveys, at redshift $z\sim0.08$ \cite{Anderson2018} with Parkes telescope data and the 2dF galaxy survey \cite{Colless2001}, and another at $z\sim0.8$ with Green Bank Telescope (GBT) data \cite{Masui2013} and the WiggleZ survey \cite{Parkinson2012}.  Cross-correlation is robust against foreground contamination because galaxies and HI trace the same underlying large-scale structure, while the foregrounds do not.  

Even when the signal can be detected it is challenging to interpret.  Intensity maps are typically analyzed using power spectra.  A density field is fully described by its power spectrum if that field is Gaussian.  However, 21 cm maps weight galaxies by their gas content, which is determined by complex, nonlinear baryon dynamics within dark matter halos.  As a result, the resulting intensity field is non-Gaussian.

For example, consider an experiment like Parkes/2dF or GBT/WiggleZ.  At $z<1$, virtually all of the neutral hydrogen is found within halos, and we can describe how it is distributed with the HI mass function (HIMF) $\phi_{\rm{HI}}(M_{\rm{HI}})$, which gives the number density of halos with HI masses between $M_{\rm{HI}}$ and $M_{\rm{HI}}+dM_{\rm{HI}}$.  Let us consider an HIMF with a slightly-modified Schechter form:
\be
\phi(\MHI)=\ln(10)\phi_*\left(\frac{\MHI}{M_*}\right)^{1+\alpha}e^{-\MHI/M_*-M_{\rm{min}}/\MHI}.
\label{HIMF}
\ee
This basic form has been used many times in the literature \cite{Schechter1976,Zwaan2005,Jones2018}. We have added an additional low-mass cutoff at $M_{\rm{min}}$, necessary as intensity maps lack hard detection thresholds.  A power spectrum, sensitive to only the Gaussian part of a field, can only access the first two moments of the HIMF \cite{Lidz2011}, while it would take at least four numbers, $\left(\phi_*,M_*,\alpha,M_{\rm{min}}\right)$, to fully determine Eq. (\ref{HIMF}).  Furthermore, there are substantial degeneracies between $\phi$ and fundamental cosmological parameters that intensity maps might otherwise be ideally situated to measure, such as halo bias, the growth rate of density fluctuations, or the amplitude of primordial non-Gaussianity \cite{Obuljen2018,Chen2018,Castorina2019}.

It was suggested in Refs. \cite{Breysse2016c,Breysse2016} that this non-Gaussianity could be accessed using one-point statistics, as opposed to two-point statistics like power spectra.  A technique called P(D) analysis, which has seen use for decades across the electromagnetic spectrum \cite{Scheuer1957,Barcons1994,Windridge2000,Lee2009,Patanchon2009,Glenn2010}, allows mapping between the HIMF and the probability distribution function (PDF) $\p(T)$ of a voxel (or three-dimensional pixel) having intensity between $T$ and $T+dT$.  This statistic, termed the Voxel Intensity Distribution (VID), has been shown to add significantly to the astrophysical information which can be gained from an intensity map \cite{Ihle2019}.

Unfortunately, as with the auto-spectrum, the VID of a 21 cm map would be dangerously contaminated with foregrounds.  Even after foreground cleaning, there can be substantial residual contamination.  Rather than expose ourselves to this large potential source of bias, we introduce here a one-point analogue to the cross-spectrum, extending the VID formalism to remove foreground bias.  We will then forecast how this technique can be applied to the Parkes/2dF and GBT/WiggleZ data.

The approach we propose here relies on a simple  fact: for independent random variables $T_1$ and $T_2$, the PDF $\p_{1+2}(T)$ of their sum $T=T_1+T_2$ is the \emph{convolution} of their individual PDFs  $\p_1(T)$ and $\p_2(T)$.  This is straightforward to prove.  We can write
\be
\p_{1+2}(T) = \int\int \p_1(T')\p_2(T'')\delta_D(T-T'-T'')dT'dT'',
\ee
where $\delta_D$ is a Dirac delta function.  Evaluating one integral leaves a convolution,
\be
\p_{1+2}(T)=\int\p_1(T')\p_2(T-T')dT'=(\p_1\circ\p_2)(T),
\label{sumconv}
\ee
We can further make use of the Fourier convolution theorem to turn this into a product in Fourier space:
\be
\fp_{1+2}(\mathcal{T})=\fp_1(\mathcal{T})\fp_2(\mathcal{T}),
\label{FTconv}
\ee
where $\T\equiv2\pi/T$ is the Fourier conjugate of the intensity $T$ and $\fp(\T)$ is the Fourier transform of $\p(T)$ (also known as the characteristic function).  In an intensity map, the observed $T$ in a given voxel is the sum of a signal component $T_S$ with PDF $\p_S$ and a contribution from noise and foregrounds which we will abbreviate as $T_{\rm{FG}}$ with PDF $\p_{\rm{FG}}$.  The full VID of a map is then the convolution of $\p_S$ and $\p_{\rm{FG}}$.

As stated above, $\p_{\rm{FG}}$ is hard to model to sufficient precision, so it will be difficult to apply the VID statistic directly to 21 cm data. We will therefore use a separate data set with different systematics to isolate our signal.  Assume that our volume contains both a 21 cm map and an optical galaxy survey.  In each voxel we know the total radio intensity $T_S+T_{FG}$ and the number $N_{\rm{det}}$ of detected optical galaxies. $T_S$ and $\Nopt$ will be correlated, due both to large-scale structure and the HI content of the optical galaxies.  We can therefore construct \emph{conditional} PDFs $\p(T|\Nopt)$.  We refer to these PDFs as conditional Voxel Intensity Distributions, or CVIDs.

Crucially, each CVID will be a convolution of a signal part, which depends on $\Nopt$, and a noise/foreground part, which does not.  If we compare voxels with different $\Nopt$, we can write,
\be
\frac{\fp(\T|\Nopt^1)}{\fp(\T|\Nopt^2)}=\frac{\fp_S(\T|\Nopt^1)\fp_{\rm{FG}}(\T)}{\fp_S(\T|\Nopt^2)\fp_{\rm{FG}}(\T)}=\frac{\fp_S(\T|\Nopt^1)}{\fp_S(\T|\Nopt^2)}.
\label{cvid_continuous}
\ee
From the second equality, it is clear that the above ratio is \emph{unbiased} by foregrounds, as the deconvolution cancels out the component which is common to both CVIDs.

In practice, we do not compute continuous PDFs directly from maps.  Instead, we estimate PDFs using histograms $B_i\approx\p(T_i)\Delta TN_{\rm{vox}}$, where $B_i$ is the number of voxels in a bin of width $\Delta T$ centered at $T_i$, and $N_{\rm{vox}}$ is the total number of voxels.  If we separate our map by $\Nopt$ values, compute histograms $B_i^{\Nopt}$ from each part, then compute their Fourier transforms $\tilde{B}_i^{\Nopt}=\Delta T\sum_j B_j^{\Nopt}\exp(i\kappa_iT_j)$, we can write down the CVID Ratio (CVR),
\be
\tilde{\mathcal{R}}^{\Nopt^1\Nopt^2}_i=\frac{\tilde{B}_i^{\Nopt^1}}{\tilde{B}_i^{\Nopt^2}}.
\label{cvr}
\ee
It is easy to show that the expectation value of $\tilde{\mathcal{R}}^{\Nopt^1\Nopt^2}_i$ is proportional to the ratio from Eq. (\ref{cvid_continuous}), and thus is unbiased by foregrounds.

\begin{figure*}
\centering
\includegraphics[width=\textwidth]{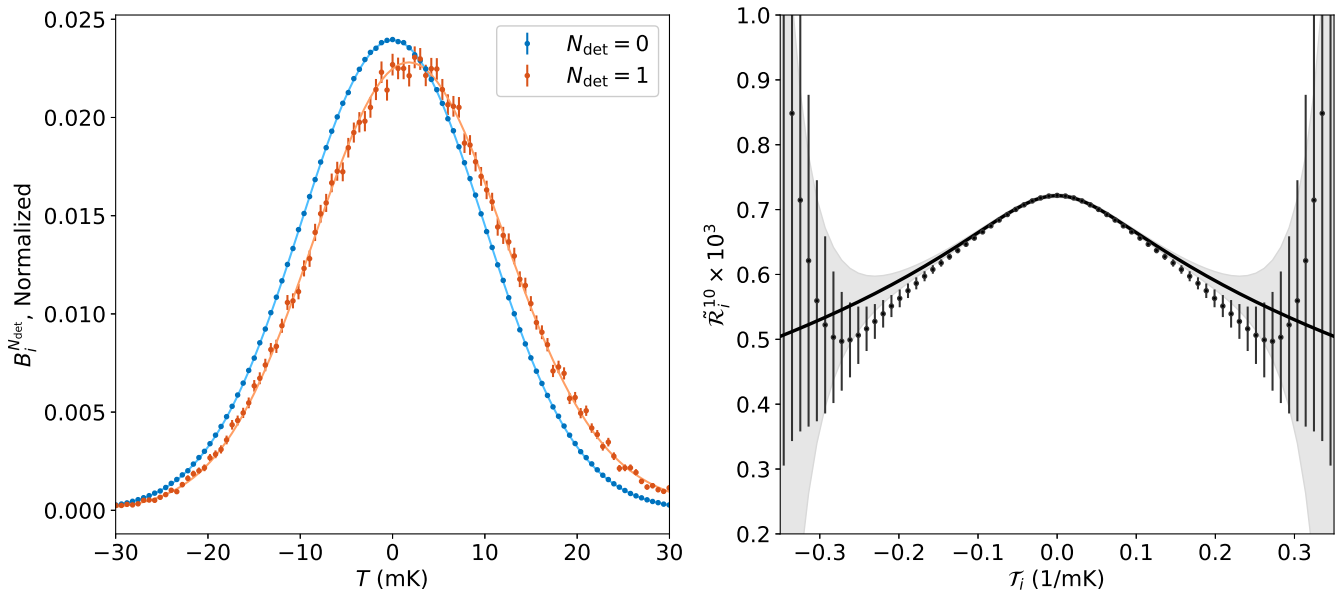}
\caption{Example CVR analysis.  (left) Example CVIDs for a model of Parkes/2dF data with 10 mK Gaussian noise, for voxels with $\Nopt=0$ (blue) and 1 (orange).  Solid curves show the theoretical expectation, points show a single realization from a toy simulation as described in the text.  Histograms are normalized to sum to unity.  See below for details of the astrophysical modeling. (right panel) CVR computed from the signal model only (solid) along with that estimated from the simulated realization.  The gray band shows the predicted $1\sigma$ error from the model.  Note that CVR errors are highly correlated, and that we plot only the real part of the (in general) complex CVR.}
\label{fig:cvr_astro}
\end{figure*}

Figure \ref{fig:cvr_astro} shows an example of how this works.  The left panel shows predicted histograms for a Parkes/2dF data set with a Gaussian $\p_{\rm{FG}}$, with the signal contribution computed as described below.  The right panel shows the CVR computed \emph{only from the signal}.  Both panels include, for illustration, a toy data realization.  If we assume that $T_S$ and $T_{\rm{FG}}$ in each voxel are independent draws from $\p_S$ and $\p_{\rm{FG}}$, then each $B_i$ is an independent draw from a binomial distribution with mean $B_i$ and variance
\be
\textrm{var}(B_i)=\left<B_i\right>(1-\left<B_i\right>/N_{\rm{vox}})\approx \left<B_i\right>.
\label{histerr}
\ee
Based on the nearly-diagonal correlation matrices shown in Figure 2 of Ref. \cite{Ihle2019}, this approximation seems reasonable, at least for low signal-to-noise data.  We can therefore create a sample histogram by drawing randomly from binomial distributions, with errors following Eq. (\ref{histerr}).  Detailed testing of this binomial approximation will require actual mock maps, which we leave for future work.

From our toy data, it is clear that the simulation, which includes foregrounds, gives the same CVR as the foreground-free theory.  Though we used a Gaussian $\p_{\rm{FG}}$, the same would hold for any general PDF.  If we can estimate the error on $B_i^{\Nopt}$, either with Eq. (\ref{histerr}) or with simulations, then we can propagate this error through to get the error on the CVR directly from the data.  This has two important implications.  First, it means that we do not need a model of the foregrounds to estimate errors on $\tilde{\mathcal{R}}^{10}_i$.  Second, it accounts for instabilities which appear when the denominator of Eq. (\ref{cvr}) approaches zero, as seen at high-$\T$ in Figure \ref{fig:cvr_astro}.  Though the measured CVR deviates significantly from the expectation, the large error bar means that these points get correspondingly little weight.

Now we need to connect the CVR to the HIMF.  We can relate the two using a modified $P(D)$ analysis \cite{Lee2009,Breysse2016,Chen2017}.  In each voxel, there are $\Nopt$ detected optical galaxies and an unknown number $N_{\rm{un}}$ of unresolved HI emitters, so:
\be
\fp(\T|\Nopt)=\fp_{\rm{det}}(\T|\Nopt)\fp_{\rm{un}}(\T|\Nopt).
\label{fpfull}
\ee
To compute $\fp_{\rm{det}}$ and $\fp_{\rm{un}}$, we need separate HIMFs for the detected and undetected galaxy populations.  For now, we assume an exponential separation,
\be
\phi_{\rm{un}}(\MHI)=\phi(\MHI)e^{-\MHI/M_{\rm{cut}}},
\label{LFun}
\ee
\be
\phi_{\rm{det}}(\MHI)=\phi(\MHI)\left(1-e^{-\MHI/M_{\rm{cut}}}\right),
\label{LFdet}
\ee
with free parameter $M_{\rm{cut}}$.

If $\Nopt=0$, there is no contribution from detected galaxies and $\p_{\rm{det}}(T|0)=\delta_D(T)$.  If $\Nopt=1$, then $\p_{\rm{det}}(T|1)$ is proportional to $\phi_{\rm{det}}$, with appropriate normalization (see \cite{Breysse2016} for details). For higher values of $\Nopt$, we can recursively apply Eq. (\ref{FTconv}) to get
\be
\fp_{\rm{det}}(\T|\Nopt)=\left[\fp_{\rm{det}}(\T|\Nopt=1)\right]^{\Nopt}.
\label{fpdet}
\ee
Note that in Eq. (\ref{fpdet}) we have implicitly assumed that the HIMF in a voxel is independent of how many galaxies it contains.  This is known to be inaccurate, as more massive objects will be more strongly biased, shifting the HIMF to larger masses in dense voxels.  Accurately modeling this effect will likely require simulations, so we neglect it for now.

For undetected galaxies, we do not know the value of $N_{\rm{un}}$, so we have
\be
\fp_{\rm{un}}(\T|\Nopt)=\sum_{N_{\rm{un}}}\left[\fp_{\rm{un}}(\T|N_{\rm{un}}=1)\right]^{N_{\rm{un}}}\p(N_{\rm{un}}|\Nopt).
\label{fpun}
\ee
where $\p_{\rm{un}}(\T|N_{\rm{un}}=1)$ is proportional to $\phi_{\rm{un}}$, $\p(N_{\rm{un}}|N_{\rm{det}})$ describes the correlation between optical galaxies and unresolved HI emitters.  If there is no clustering, $\p_{\rm{un}}$ has no dependence on $\Nopt$, and will cancel out of the CVR.  However, we know that HI and optical galaxies should be at least somewhat correlated, though that correlation may be color- and scale-dependent \cite{Hess2013,Papastergis2013,Anderson2018}.  Here we will employ a simple model to capture the leading order effects.

\begin{figure*}
\centering
\includegraphics[width=\textwidth]{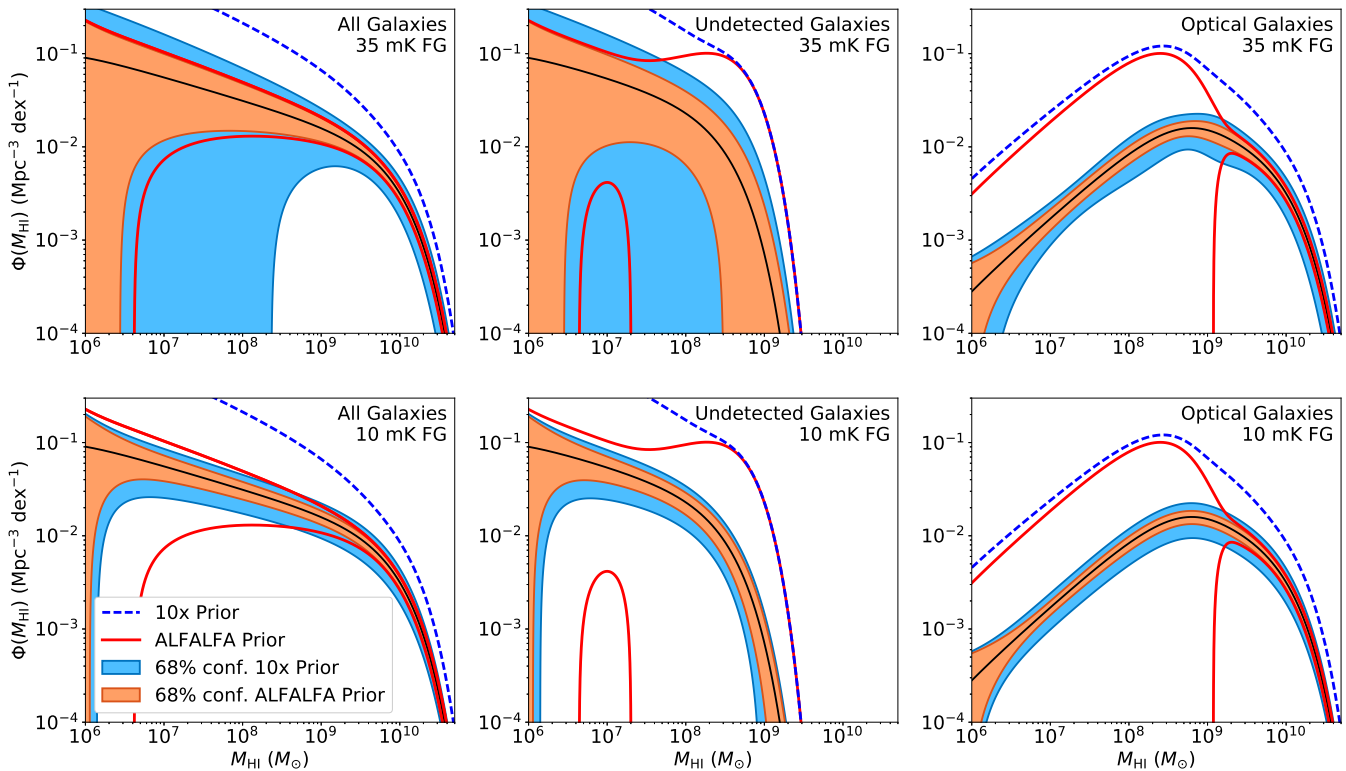}
\caption{Forecasted CVR constraints on the HIMF of all galaxies (left column), those undetected by the optical survey (center column), and those detected optically (right column). Black solid lines show our fiducial model (Eqs. (\ref{HIMF},\ref{LFun},\ref{LFdet}) with ALFALFA parameters), the other lines show $1\sigma$ prior uncertainties for the ALFALFA priors (red solid) and the 10x ALFALFA priors (blue dashed).  Shaded regions show the $1\sigma$ regions after our Fisher forecast, both with the ALFALFA (red) and 10x ALFALFA priors (blue).  The upper row shows the case with modest foreground cleaning ($\sigma_{FG}=35$ mK), the lower row with stronger foreground cleaning ($\sigma_{FG}=10$ mK).}
\label{fig:results}
\end{figure*}

Following \cite{Breysse2016,Agrawal2017}, assume each voxel has expected counts $\mu_{\rm{un}}$ and $\mu_{\rm{det}}$ of detected and undetected galaxies, and that the actual counts are Poisson draws with these means.  We take $\mu_{\rm{un}}=f_{\rm{un}}\mu_{\rm{det}}$, where $f_{\rm{un}}$ is the ratio $\overline{N}_{\rm{un}}/\overline{N}_{\rm{det}}$ of the mean number counts.  Galaxy count distributions are known to be reasonably-approximated by lognormal distributions \cite{Coles1991,Kayo2001}, 
\be
\p_{\rm{LN}}(\mu)=\frac{1}{\mu\sqrt{2\pi\sigma_G^2}}\exp\left\{-\frac{1}{2\sigma_G^2}\left[\ln\left(\frac{\mu}{\overline{N}}\right)+\frac{\sigma_G^2}{2}\right]^2\right\},
\ee
where the $\sigma_G$ parameter acts as the ``bias" in this model.  The uncorrelated case corresponds here to $\sigma_G=0$.  We then can write
\be
\p(N_{\rm{un}}|\Nopt)=\int\p(N_{\rm{un}}|f_{\rm{un}}\mu_{\rm{det}})\p(\mu_{\rm{det}}|\Nopt)d\mu_{\rm{det}}.
\ee
We can use Bayes' Theorem to state that,
\be
\p(\mu_{\rm{det}}|\Nopt)\propto\p(\Nopt|\mu_{\rm{det}})\p_{\rm{LN}}(\mu_{\rm{det}}),
\ee
where $\p_{\rm{LN}}$ acts as our ``prior".  With our assumption of Poisson statistics, we then have
\begin{multline}
\p(N_{\rm{un}}|\Nopt)\propto\int\p_{\rm{Poiss}}(N_{\rm{un}}|f_{\rm{un}}\mu_{\rm{det}})\p_{\rm{LN}}(\mu_{\rm{det}}) \\
\times \p_{\rm{Poiss}}(N_{\rm{det}}|\mu_{\rm{det}})d\mu_{\rm{det}},
\label{pundet}
\end{multline}
where $\p_{\rm{Poiss}}(N|\mu)$ is the Poisson distribution with mean $\mu$.

We can now finally predict a CVR from our HIMF model.  We will now examine what information could be gained from such an analysis.  Consider a model of the Parkes/2dF maps as described in Ref. \cite{Anderson2018}, with free parameters $(\phi_*,M_*,\alpha,M_{\rm{min}},M_{\rm{cut}},\sigma_G)$.  Assume that HI evolves negligibly from $z=0-0.05$, and use the best-fit HIMF from the ALFALFA survey \cite{Jones2018} as a model, with $\phi_*=(4.5\pm0.8)\times10^{-3}$ Mpc$^{-3}$ dex$^{-1}$, $\log(M_*/M_{\odot})=9.94\pm0.05$, and $\alpha=-1.25\pm0.1$.  We arbitrarily choose $M_{\rm{min}}=10^5\ M_{\odot}$ so that it falls below the ALFALFA detection threshold, and $M_{\rm{cut}}=3\times10^8\ M_{\odot}$ to get the correct number of optically detected galaxies \cite{Colless2001}.  We compute $\sigma_G$ following Ref. \cite{Breysse2016}.  We assume a Gaussian $\p_{\rm{FG}}$, but note again that this procedure would work regardless of the assumed form.  As the strength of the contamination depends on how much prior foreground cleaning is assumed, we consider a pessimistic case with $\sigma_{\rm{FG}}=35$ mK and an optimistic case with $\sigma_N=10$ mK, arbitrarily chosen to very roughly match the Parkes maps from Figure 1 of Ref. \cite{Anderson2018}.  Note that we do not allow $\phi_*$ and $\alpha$ to vary between the detected and undetected populations, but assume a single value for each.

We can forecast constraints on our model using the Fisher matrix formalism \cite{Fisher1935,Tegmark1997}.  We only consider $\Nopt=0$ and 1 here, as few data voxels have $N_{\rm{det}}>1$.  As we neglect evolution from the $z=0$ ALFALFA galaxies, we can use their quoted systematic errors as priors on $\phi_*$, $M_*$ and $\alpha$.  Though we do not directly forecast the higher-redshift GBT/WiggleZ survey, we can get a rough idea of its performance using a weaker set of priors, which we take as ten times worse than the ALFALFA priors.  In both cases, we assume $10\%$ prior knowledge of $\sigma_G$, which would have to come from simulations, and uninformative fractional priors of 10 on both $M_{\rm{cut}}$ and $M_{\rm{min}}$.  We assume a cosmology consistent with the Planck 2015 results \cite{Planck2016}.

Figure \ref{fig:results} shows the results of our Fisher forecasts.  We plot $1\sigma$ confidence intervals around our fiducial models for the total, detected, and undetected populations.  Even in the most pessimistic case, with strong foregrounds and weak (10x ALFALFA) priors, we get a good measurement of the bright end of the HIMF and the detected galaxy HIMF.  If we know enough to trust the strong ALFALFA priors, then the CVID adds important constraints on the HIMF of the optical galaxies and a modest measurement of that of the unresolved galaxies, both of which cannot be obtained from ALFALFA alone.  With 10 mK foregrounds, the CVID dramatically improves on the priors.  The very brightest end of the HIMF is still dominated by ALFALFA, but the CVID has added a wealth of information about faint galaxies which cannot be obtained conventionally.  Unfortunately, even the intensity mapping data loses sensitivity at the very brightest end of the HIMF.  Even stronger foreground cleaning would be needed to measure $M_{\rm{min}}$.

These forecasts clearly demonstrate the utility of this method.  However, we have made a number of assumptions here that deserve further study.  The lognormal form of the galaxy number count PDF is likely overly simplistic, and could be replaced by a more sophisticated prescription \cite{Leicht2019}.  In Eqs. (\ref{LFun}) and (\ref{LFdet}), we assumed that optical galaxies host the brightest 21 cm emitters.  This should hold for blue, gas-rich galaxies, but a number of bright optical galaxies are red and gas-poor \cite{Cole2005}, and therefore have weaker HI emission.  As mentioned above, we have entirely neglected luminosity-dependent bias in our forecasting.  Though the Parkes and GBT surveys are single-dish, many 21 cm experiments are interferometric, which may introduce extra systematics.  Finally, we assumed that we can apply some degree of foreground cleaning to our data without affecting our signal.  In the Parkes foreground cleaning, the signal was suppressed along with the foregrounds \cite{Anderson2018}, which if uncorrected would bias our CVR measurements.  These caveats motivate additional study of this method using mock data sets, as in Ref. \cite{Ihle2019}.  

Though we have focused here on the HIMF, the potential utility of this method extends much further.  For example, one could study the HI-color relation mentioned above \cite{Wolz2017}.  With intensity maps of other lines \cite{Kovetz2019,Crites2014,Keating2016,Li2016,Gong2017,Croft2018,Origins2018,Padmanabhan2018,Stacey2018,Cooray2019,Dumitru2019}, one could, for example, measure the molecular gas properties of Lyman-alpha emitters \cite{Chung2019}, or study AGN feedback \cite{Breysse2019}.  The CVR can also be modified to combine intensity maps of different lines, another common experimental target \cite{Lidz2011,Breysse2016b,Fonseca2018,Beane2019}.

Cross-correlations have long been a powerful tool for cosmologists, and will only become more critical as more intensity mapping surveys come online.  With this work we have demonstrated a one-point cross-correlation method that can be used to clean foregrounds and probe astrophysics inaccessible to conventional surveys.  With some refinement, this will be a valuable tool for many future experiments.

The authors would like to thank Ue-Li Pen and Hamsa Padmanabhan for useful conversations.  Philippe Berger was supported by Jet Propulsion Laboratory, California Institute of Technology, under a contract with the National Aeronautics and Space Administration. Copyright 2019. All rights reserved.


\end{document}